# Dynamics of single-photon emission from electrically pumped color centers

I. A. Khramtsov,[1] M. Agio,[2,3] and D. Yu. Fedyanin[1,*]

[1] *Laboratory of Nanooptics and Plasmonics, Moscow Institute of Physics and Technology, 141700 Dolgoprudny, Russian Federation*
[2] *Laboratory of Nano-Optics, University of Siegen, 57072 Siegen, Germany*
[3] *National Institute of Optics (CNR-INO) and Center for Quantum Science and Technology in Arcetri (QSTAR), 50125 Florence, Italy*

*email: dmitry.fedyanin@phystech.edu

**Abstract**

Low-power, high-speed and bright electrically driven true single-photon sources, which are able to operate at room temperature, are vital for the practical realization of quantum communication networks and optical quantum computations. Color centers in semiconductors are currently the best candidates, however, in spite of their intensive study in the past decade, the behavior of color centers in electrically controlled systems is poorly understood. Here we present a physical model and establish a theoretical approach to address single-photon emission dynamics of electrically pumped color centers, which interprets experimental results. We support our analysis with self-consistent numerical simulations of a single-photon emitting diode based on a single nitrogen-vacancy center in diamond and predict the second-order autocorrelation function and other emission characteristics. Our theoretical findings demonstrate remarkable agreement with the experimental results and pave the way to the understanding of single-electron/single-photon processes in semiconductors.

## I. INTRODUCTION

Recent developments in quantum optical technologies hold promise for realizing unconditionally secure communication lines based on quantum cryptography [1], precision measurements below the shot-noise limit [2] and optical quantum computers, which are able to do things that cannot be done by any machine based on classical physics [3]. An essential element of these devices is a generator of non-classical light, namely a single-photon source, which produces a train of optical pulses so that each of them contains one and only one photon. However, obtaining of a bright, stable and efficient true single-photon source suitable for practical applications is still a great challenge for quantum optoelectronics. Such sources should be free from blinking, have a narrow emission spectrum, be integrable into large-scale quantum circuits [4,5] and operate at room temperature [6–8]. Implementation of electrical pumping is also strongly needed, since optically pumped single-photon sources are characterized by low energy efficiency, poor scalability and require bulky external high-power pump optical sources [7]. Some defects, known as color centers, in the crystal lattice of diamond and other wide bandgap semiconductor materials, such as silicon carbide, gallium nitride, zinc oxide and hexagonal boron nitride, are considered to be the most promising candidates for the role of electrically driven single-photon sources [8]. However, to date, electroluminescence from only a few defects in diamond and silicon carbide has been observed [9–14], the physics behind the process of single-photon emission under electrical pumping is not yet well understood.

In this work, we introduce a theoretical description and present an interpretation of the recent experimental results for the single-photon emission dynamics of color centers under electrical pumping. We also provide a comprehensive study of NV centers in a diamond p-i-n diode and present our findings in a ready-to-use form, which can be applied to any color center in any semiconductor material. In addition, our work can be extended to study single-photon emission dynamics of electrically pumped epitaxial quantum dots, for which a theoretical calculation of the second-order correlation function has not been reported yet. We support our theoretical results with rigorous self-consistent simulations of the diamond single-photon emitting diodes and reproduce experimental results.

## II. RESULTS AND DISCUSSION

### A. Single-photon emission dynamics in an electrically pumped system

The process of single-photon emission from color centers under electrical pumping is much more complicated than the process of photon emission under optical pumping. For example, the recently measured second-order autocorrelation function of electrical pumped color centers in diamond and silicon carbide demonstrates a triple exponential dependence [9], which is remarkably different from the two exponential dependence observed for these centers under optical pumping. One of the hypothesis was the existence of an additional level in the energy spectrum of the neutrally charged NV center [10]. However, here we demonstrate that this sophisticated state is, in fact, responsible for the ground state of the negatively charged NV center rather than for the excited state of the neutral NV center. We also show that there are no direct electron transitions among the states of the color center except the radiative transition and additional electron transitions between the conduction and valence bands of the semiconductor and electron (or hole) levels of the color center must be considered.

Color center excitation by optical pumping involves only transitions between the ground and excited states of the center (here we do not consider optical pumping at very high photon energies and two-photon absorption at high pump rates, which can lead to color center ionization). In this case, there is no interaction between the color center and the semiconductor crystal: the defect in the lattice structure behaves like an isolated atom except interaction with phonons, which only broadens the emission spectrum [15]. In contrast to the optical excitation, electrical pumping implies electron exchange between the color center and the semiconductor via carrier capture and release [16]. Free excitons may also contribute to the photon emission from the color center. However, it is known that in semiconductors, at moderate and high injection levels, the recombination processes at the defects are not noticeably affected by the excitonic effects even at liquid nitrogen temperatures [17]. Our estimations of the free exciton density at room temperature in diamond [18,19] show that it is at least 4 orders of magnitude lower than the density of electrons, while the capture cross-sections and thermal velocities are of the same order of magnitude [20–23]. Thus, even at high pump currents, free excitons cannot noticeably affect the emission characteristics.

The rate and the speed of the electron and hole exchange between the color center and the conduction and valence bands of the semiconductor along with the internal properties of the color center determine characteristics of the electrically pumped single-photon source. While the brightness of a single emitter can be measured as the output power in a steady state regime, many dynamic characteristics can be extracted by measuring the quantum correlation among emitted photons. When the color center emits a photon at time $\tau = 0$, it necessarily appears in the ground state, from which it cannot emit photons. After that, it takes a nonzero time to "recharge" the color center, which can be clearly observed in the experimentally measured second-order photon autocorrelation function $g^{(2)}(\tau)$ [9,10,13] [Fig. 1(a)]. Since there is a dead time between photon emission events, $g^{(2)}(0) = 0$. At the same time $g^{(2)}(\tau \to \infty) = 1$ due to the absence of correlation between photons at large delay times. The shape and characteristic times of the $g^{(2)}$ curve can give information about the speed of the transitions among energy states of the color center and about the speed and intensity of the electron and hole exchange between the color center and the semiconductor crystal. Unfortunately, this cannot be done straight forward, and deep understanding of physics behind the electroluminescence process is needed to interpret the experimental results and predict properties of electrically driven color centers.

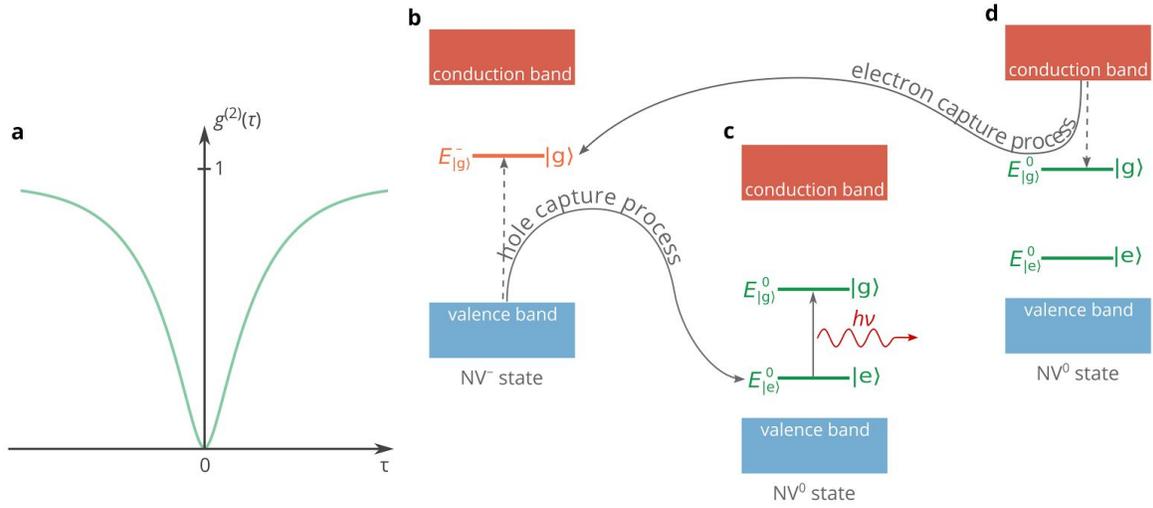

**FIG. 1.** (a) Schematic illustration of the typical $g^{(2)}$ function of electrically pumped color center. (b-d) Diagram of the three-stage process of single-photon emission from the NV center in diamond under electrical pumping: (b) hole capture from the valence band, (c) photon emission, (d) electron capture from the conduction band (or hole release to the conduction band). In panels (b) and (c), the excited level $E^0_{|e\rangle}$ is shown below the ground level $E^0_{|g\rangle}$, since transitions in the NV$^0$ center are related to the capture of a hole and its subsequent relaxation to the ground level. This energy level is physically responsible for the known $^2A_1$ excited state of the NV$^0$ center.

In this work, we focus on nitrogen-vacancy (NV) centers in diamond, which are the most studied color centers in semiconductors to date [8,15]. At the same time, we note that our results can be easily generalized and applied to any color center in any semiconductor, since the electron and hole exchange rates are mostly determined by the properties of the semiconductor and the charge of the defect, rather than by the internal structure of the defect, which is shown below. The NV center in diamond has two charge states: negatively-charged (NV$^-$) and neutral (NV$^0$). It has been demonstrated that under optical excitation both charge states (NV$^-$ and NV$^0$) can emit photons under optical pumping [15,24,25]. Under electrical pumping, photon emission only from the neutral NV$^0$ center was observed [10,13]. The reason for this is that the color center can be excited only by means of electron and hole exchange between the color center and the semiconductor crystal, since this is the only way to transfer a significant amount of energy to a defect in the crystal lattice [16]. We limit our consideration to a simplified two-level model of the NV$^0$ center, while a more complicated three-level model will be discussed in the next section. Figure 1(b-c) shows the three-stage process of photon emission under electrical pumping. In the first stage, the negatively charged color center attracts positively charged holes and can capture them. The hole capture process changes the charge of the color center. As a result, the center appears in the neutrally charged NV$^0$ excited state. In the second stage, the NV$^0$ excited state spontaneously relaxes to the NV$^0$ ground state [Fig. 1(c)] either radiatively or not. Let us introduce the quantum efficiency of this transition $\eta$, which is in the range from 30% to 80% [10,25]. Since the experimental studies of NV$^0$ centers do not reveal noticeable photon bunching, we can easily find the lifetime of the excited state as $\tau_0 = \eta \tau_r$, where $1/\tau_r$ is the radiative transition rate. This approach gives the possibility to take into account nonradiative transitions using a simple two-level model. In the third stage [see Fig. 1(d)], the NV center in the NV⁰ ground state could capture a negatively charged electron into the NV⁻ excited state. However, the experimental studies do not show emission from this state under electrical pumping [10,13,14]. Our analysis of the color center as a multielectron system [26,27] demonstrates that an electron cannot be captured into the NV⁻ excited state, since the maximum energy which the color center can get by capturing an electron is lower than the difference between energies of the NV⁰ ground state and NV⁻ excited state calculated as the sum of energies of the electrons of the color center. Therefore, the electron capture transforms the NV⁰ ground state into the NV⁻ ground state [see Fig. 1(d,a)].

Following Fedyanin and Rzhanov [16,28], we can address quantitatively the dynamics of the process of single-photon emission from a single color center under electrical pumping:

$$\begin{cases} \frac{dx}{dt} = c_p p(1-x-f) - e_p x + e_r f - x/\tau_0 \\ \frac{df}{dt} = e_n(1-x-f) - c_n n f + x/\tau_0 - e_r f \end{cases} \quad (1)$$

This is the system of rate equations for the populations $x$, $f$ and $(1-x-f)$ of the NV⁰ excited state, NV⁰ ground state, and NV⁻ ground state, respectively. Here, $c_n = \sigma_n \langle v_n \rangle$ and $c_p = \sigma_p \langle v_p \rangle$ are the electron and hole capture rate constants ($\sigma_n$ and $\sigma_p$ are the carrier capture cross-section and $\langle v_n \rangle$ and $\langle v_p \rangle$ are the average carrier thermal velocities in the semiconductor); $e_n$ and $e_p$ are the electron and hole emission constants, which show the probability of thermal emission of electrons to the conduction band ($e_n$) and holes to the valence band ($e_p$) of the semiconductor; $e_r$ characterizes thermal excitation from the NV⁰ ground state to the NV⁰ excited state. The electron capture cross-section by the NV⁰ center is roughly equal to $10^{-15}$ cm² [22], while the hole capture cross-section by the NV⁻ center is calculated using the cascade capture theory [16,17] to be $3.2 \times 10^{-14}$ cm² at room temperature. For details on the calculations of $e_n$ and $e_p$, see Ref. [17]. The g⁽²⁾ function can be found by solving the system of equation (1) with the initial conditions $x=0$ and $f=1$ [29] at $\tau=0$, which correspond to the color center in the NV⁰ ground state right after photon emission. Eventually, one can obtain an analytical expression for the g⁽²⁾ function:

$$g^{(2)}(\tau) = 1 + a e^{-|\tau|/\tau_1} - (1+a) e^{-|\tau|/\tau_2}, \quad (2)$$

where

$$\tau_{1,2} = 2\left\{ \left[1/\tau_0 + c_n n + c_p p + (e_r + e_n + e_p)\right] \pm \right.$$
$$\left. \pm \sqrt{\left[c_n n - c_p p - 1/\tau_0 + (e_n + e_r - e_p)\right]^2 - 4\left[c_p p/\tau_0 - (c_p p e_n + e_r/\tau_0 - e_n e_r)\right]} \right\}^{-1} \quad (3)$$

and

$$a = \frac{1}{\tau_2 - \tau_1}\left[\tau_1 - \frac{e_r}{c_n n c_p p + (c_p p + e_n)e_r}\right] \quad (4)$$

In general, both $\tau_1$ and $\tau_2$ can be complex, so can *a*. At the same time, we emphasize that the g$^{(2)}$ function always takes only real values. Equation (2) has the same form as the expression for the g$^{(2)}$ function of the optically pumped three-level system (e.g., see Refs. [24,29,30]). It consists of two exponential terms: one is responsible for bunching and one for antibunching. However, the nature of these exponential terms is different: they include the electron and hole exchange rates between the color center and the conduction and valence bands of the semiconductor, as can be seen from equations (3) and (4). Thus, the kinetics of single-photon emission under electrical pumping is determined by the electron and hole densities in the vicinity of the color center [Fig. 2]. The real parts of $\tau_1$ and $\tau_2$ are responsible for the bunching and antibunching times, while the imaginary parts along with the real parts determine the ratio between the bunching and antibunching terms in the g$^{(2)}$ function. Interesting is that both Re($\tau_1$) and Re($\tau_2$) can be lower than the lifetime $\tau_0$ of the excited state of the NV$^0$ center at very high injection levels. For example, at carrier densities *n*=*p*=10$^{17}$ cm$^{-3}$, $\tau_1$ = 26 ps and $\tau_2$ = 0.5 ns and the lifetime of the excited state $\tau_0$ does not affect the emission dynamics of the color center. At the same time, $\tau_0$ apparently limits the photon emission rate.

At room temperature at a photon count rate greater than ~1 count/s, the electron and hole capture processes dominate over thermal excitations and the *e*-terms in equations (3) and (4) can be set to zero, which significantly simplifies the analysis. We can also note that, for the NV$^0$ center, $\tau_0$ is equal to 5.1 ns at a quantum efficiency of 30% [25], which is significantly smaller than the inverse electron and hole capture rates that can be achieved at room temperature [16]. Accordingly, we can derive that $\tau_1 \approx \mathrm{Re}(\tau_1) \approx \tau_0$, $\tau_2 \approx \mathrm{Re}(\tau_2) \approx 1/(c_\mathrm{n} n + c_\mathrm{p} p) \gg \tau_1$ and, as follows from equation (2), $g^{(2)}(\tau) \approx 1 - \exp(-|\tau|/\tau_2)$. The meaning of this simple expression is that the characteristic time of the g$^{(2)}$ function steadily decreases as the densities of electrons and holes increase in the vicinity of the color center, namely as the pump current increases. This is exactly what was observed experimentally in diamond [10] and silicon carbide [9] single-photon emitting diodes. At very high injection levels, which are hardly achievable at room temperature but are possible at 400-600 K [16], one should use more complicated equation (2) with complex characteristic times $\tau_1$ and $\tau_2$, the physical interpretation of those is not as evident as at low injection levels.

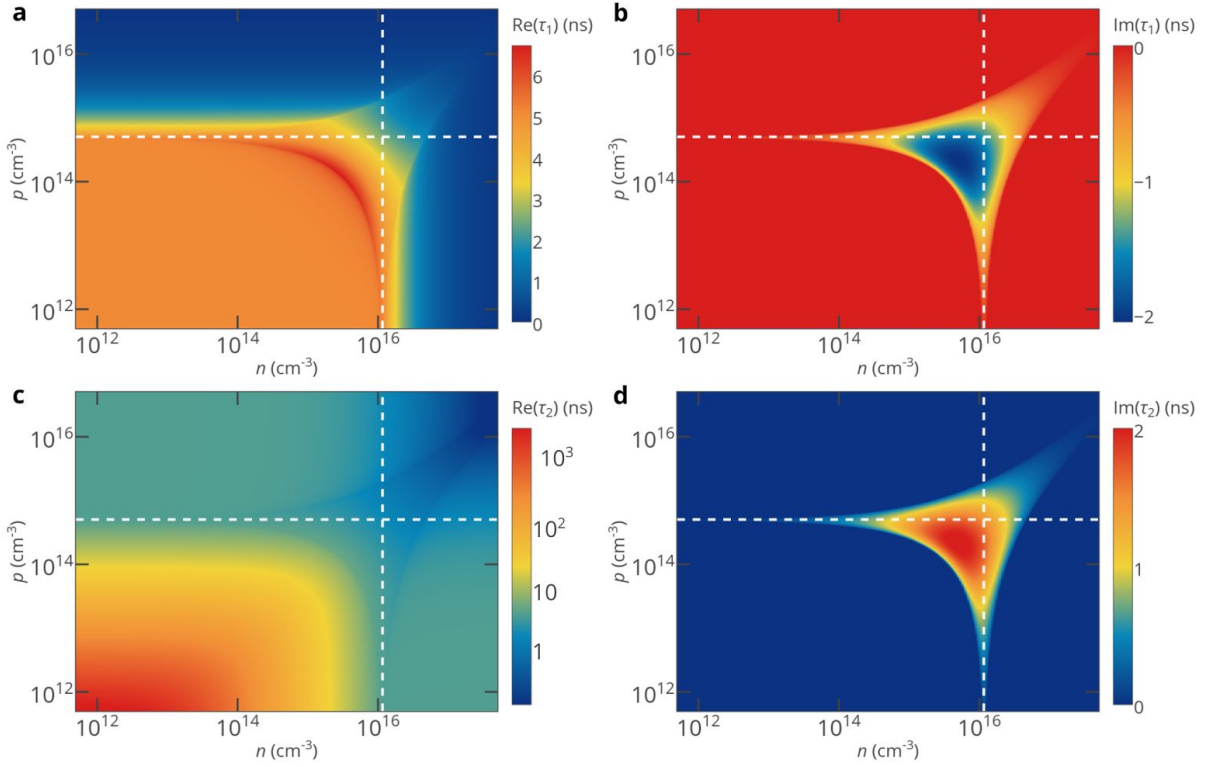

**FIG. 2.** Characteristic times $\tau_1$ (panels a and b) and $\tau_2$ (panels c and d) of the $g^{(2)}$ function for the electrically pumped NV center in diamond at room temperature, $\eta$ = 30% [25]. At dashed lines, either $c_p p = 1/\tau_0$ or $c_n n = 1/\tau_0$.

## B. Single-photon emitting diode based on a nitrogen-vacancy center in diamond

To validate our theoretical framework and interpret recently obtained experimental results, we proceed with the analysis of the single-photon emitting diode shown in Fig. 3a. The NV center is incorporated into the intrinsic region of the diamond p-i-n diode, which facilitates the delivery of charge carriers to the defect for bright electroluminescence under forward bias. This geometry was recently used by Mizuochi et al. [10] to study the evolution of the $g^{(2)}$ function under increasing pump current. The position of the NV center is dictated by the necessity to collect the emitted photon with an oil immersion objective. The p-layer is doped with boron at a concentration of $10^{19}$ cm$^{-3}$, while the concentration of phosphorus in the 500-nm-thin n-type layer equals $10^{18}$ cm$^{-3}$ [10]. The acceptor compensation ratio is set to 1%, which is typical for p-type diamond samples [31]. The donor compensation effects in the n-layer are much stronger [18,32] and the compensation ratio is assumed to be at least 10% based on the previous experimental studies of similar samples [33,34]. The experimentally measured electron mobility in the n-layer and hole mobility in p-layer are equal to 150 cm$^2$/Vs and 10 cm$^2$/Vs, respectively [10]. The carrier mobilities in the intrinsic region are much higher (~2500 cm$^2$/Vs for electrons and ~1200 cm$^2$/Vs for holes [22]) and their values do not significantly affect the results. To analyze the p-i-n diode we have performed self-consistent numerical simulations of the electron and hole transport based on the Poisson equation, drift-diffusion current equations and the electron and hole continuity equations using the nextnano software [35]. Here, we note that due to the experimental uncertainties [10], the vertical device geometry and particular features of the p-i-n structure, one dimensional and three-dimensional simulations give roughly the same results when we consider the dependencies versus the current density in the vicinity of the color center [36,37].

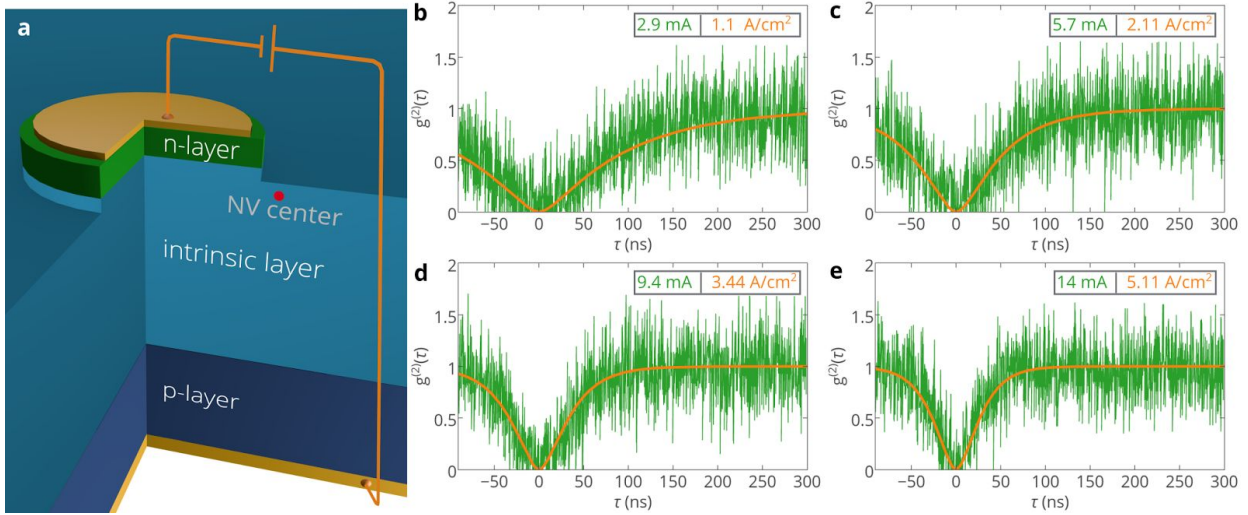

**FIG. 3.** (a) Schematic view of the single-photon emitting p-i-n diamond diode (not to scale). The diameter of the diamond disc is equal to 220 μm, the thickness of the n-type layer is 500 nm, the intrinsic region is 10 μm thick. The NV center is at a depth of 300 nm and at a distance of about 30 μm from the n-type contact. (b-e) Background corrected experimental $g^{(2)}$ curves extracted from [10] (shown in green) and numerically simulated $g^{(2)}$ functions (shown in orange) for four different pump currents. The experimentally measured current and corresponding local current density obtained in the simulations are shown in each panel. The quantum efficiency is extracted from the experiment and is equal to 78% in the simulations.

Figure 3 shows the numerically simulated and experimentally measured $g^{(2)}$ curves at four different pump currents. The $g^{(2)}$ function experimentally measured at 2.9 mA is achieved at a slightly lower current [Fig. 3(b)], since the exact value of the donor compensation ratio in the n-type layer and mobilities of minority carriers in all three layers are not known. Nevertheless, we observe perfect agreement between theoretical and experimental $g^{(2)}$ curves as the injection current increases 2 folds [Fig. 3(c)], 3.2 folds [Fig. 3(d)] and 4.8 folds [Fig. 3(e)]. Alternatively, it was possible to vary the unknown parameters and get nearly the same coincidence of experimental and theoretical curves at the same injection current, which is a consequence of the fact that the photon emission rate is determined by the densities of the electrons and holes in the vicinity of the color center rather than by the pump current and the required densities are achieved at different currents in different conditions, e.g. the current should be higher for a higher compensation ratio in the n-type layer.

It can be clearly seen that, in line with our theoretical predictions discussed above, the characteristic time of the g$^{(2)}$ function monotonically decreases as the pump current increases due to the increase in the density of nonequilibrium electrons and holes in the i-type region in the vicinity of the color center [Fig. 4(a)]. Since the electron capture cross-section by a neutral NV$^0$ center is significantly smaller than the capture cross-section of a positively charged hole by a negatively charged NV$^-$ center, the characteristic time $\tau_2$ is approximately equal to the inverse hole capture rate $1/c_p p$, which is roughly inversely proportional to the pump current [Fig. 4(a). In turn, $\tau_1$ almost does not depend on the pump current in the studied range of the pump levels and can be considered to be equal to the lifetime of the excited state. Figure 4(b) presents the evolution of the g$^{(2)}$ function with increasing pump current. At high injection currents, the single-photon emission process is orders of magnitude faster than at low pump levels, which is clearly seen visually, and the characteristic time approaches the lifetime $\tau_0$ of the excited state as the current increases. However, it is extremely difficult to decrease the characteristic time down to $\tau_0$ due to high activation energies of dopants and donor and acceptor compensation effects in diamond [18,32], which prevents us from achieving very high densities of non-equilibrium charge carriers in the i-layer of the diamond diode at room temperature.

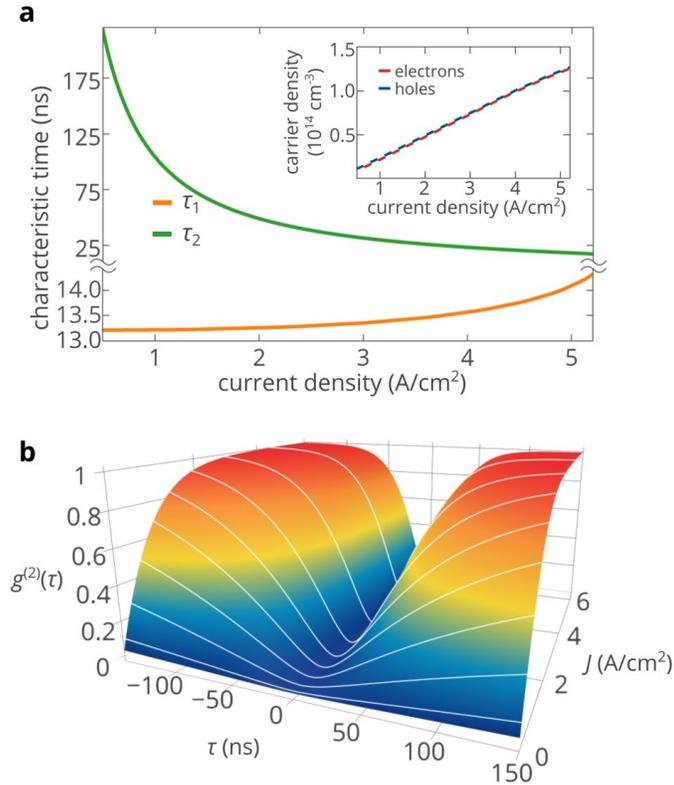

**FIG. 4.** (a) Characteristic times $\tau_1$ and $\tau_2$ of the g$^{(2)}$ function for the NV center in the p-i-n diode versus the pump current density. Inset: electron and hole densities in the vicinity of the color center as a function of the injection current. (b) Simulated evolution of the g$^{(2)}$ function with the increasing pump current.

It is important to note that our simulations do not show the presence of noticeable photon bunching in the emission from the electrically pumped NV center in diamond, which agrees well with the experimental studies [10,13,14]. The detailed analysis of the characteristics times predicts the maximum values of the $g^{(2)}$ function of only 1.004 even at very high densities of electrons and holes in the vicinity of the NV center [Fig. 2]. At the same time, it is known that the photoluminescence of the NV center is better described by a three-level model [24] rather than the two-level model with a non-unity quantum efficiency used above. The lifetime $\tau_s$ of the shelving state is not known for the $NV^0$ center, but, since photon bunching was not observed in either photoluminescence [10,24] or electroluminescence [10,13] measurements, $\tau_s$ must not exceed a few nanoseconds. To address the impact of the shelving state on the characteristics of the photon autocorrelation function, we have implemented this model in our simulations [see Fig. 5(a)]. Since the analytical expression in this case is very complicated, we have used the half-rise time $\tau_{1/2}$ of the $g^{(2)}$ function (i.e. $g^{(2)}(\tau_{1/2}) = 1/2$) to compare different physical models. Figure 5(b) clearly shows that there is no significant difference between the two- and three-level models at the experimentally achievable densities of electrons and holes. Was $\tau_s$ longer, it would produce stronger impact, and this is what can be expected for electrically pumped SiV centers in diamond [30]. For electrically pumped NV centers, the two-level model with a non-unity quantum yield can be used.

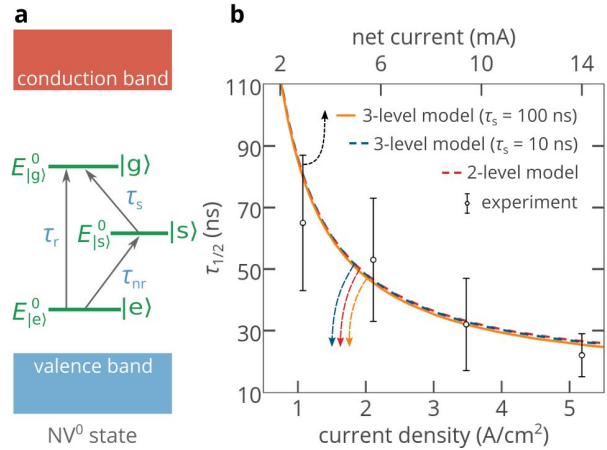

**FIG. 5.** (a) Schematic illustration of the three-level model of the $NV^0$ center, which should be used at the second stage [see Fig. 1(c)] of the process of single-photon emission under electrical pumping [Fig. 1(b-d)]. $\tau_{nr}$ is the non-radiative lifetime of the excited state and $\tau_s$ is the lifetime of the shelving state. (b) Half-rise time $\tau_{1/2}$ of the $g^{(2)}$ function versus the injection current in the diamond p-i-n diode [see Fig. 3(a)] retrieved from the experiment and obtained in the numerical simulations using either the two-level model or the three-level model, or the three-level mode with a ten times longer lifetime of the shelving state.

## III. CONCLUSIONS

We have introduced a theoretical approach for addressing single-photon emission dynamics of electrically pumped color centers and validated it by reproducing the experimental $g^{(2)}$ functions for the NV center in a diamond diode. In accordance with the proposed mechanism, the characteristic time of the $g^{(2)}$ function is determined by the electron and hole capture rates by the color center and is a function of the electron and hole densities in the vicinity of the color center. While the photon emission rate is determined by the slowest process between electron capture and hole capture processes, the emission dynamics and characteristic time of the $g^{(2)}$ function are governed by the fastest process. In the case of NV centers in diamond, the characteristic time of the $g^{(2)}$ function is determined by the fast hole capture process rather than by the slow electron capture process. Interesting is that the emission dynamics is not limited by the lifetime of the excited state of the color center and is governed only by the electron and hole capture processes at both very low and very high pump currents. At very high injection levels, characteristic times $\tau_1$ and $\tau_2$ of the $g^{(2)}$ function are orders of magnitude lower than the lifetime of the excited state of the color center, while at low pump current the half-rise time $\tau_{1/2}$ of the $g^{(2)}$ function is roughly equal to the antibunching time $\tau_2$, which is in the millisecond range. It is important to note that the hole capture rate also determines the response time of the electrically pumped single-photon source. In equilibrium, the NV center is in the NV⁻ ground state in intrinsic or n-type diamond, and the time, at which a photon is emitted, is determined by the probability to capture a hole [see Fig. 1(b-d) for details]. The response time is up to two orders of magnitude faster than the "recharge" time governed by the electron capture process, which is advantageous for obtaining of electrically pumped true single-photon sources on demand driven by short electrical pulses. The developed theoretical approach can be applied with no change to electrically pumped color centers in diamond, silicon carbide, gallium nitride, zinc oxide, hexagonal boron nitride and other semiconductors. In addition, it can be extended to quantum dots in semiconductors. We believe that our findings form a solid backbone for the understanding of physics behind single-photon emission in diverse electrically driven systems and for the development of efficient single-photon sources.

## ACKNOWLEDGEMENTS

The work was supported by the Russian Foundation for Basic Research (16-37-00509-mol_a and 16-29-03432-ofi_m), by the grant of the President of the Russian Federation (MK-2602.2017.9), by the Ministry of Education and Science of the Russian